\documentclass[conference]{IEEEtran}
\IEEEoverridecommandlockouts
\usepackage{main}
\begin{document}
\newcommand{\gnba}{\ensuremath{\text{gNB}_{\textit{ag}}} }
\newcommand{\gnbi}{\ensuremath{\text{gNB}_{\textit{in}}} }
\newcommand{\ue}[2]{\ensuremath{\text{UE}_{#1}^{\text{#2}}} }
\newcommand{\gain}[2]{\ensuremath{G(#1,#2)} }
\newcommand{\power}[1]{\ensuremath{P_{#1}} }

\begin{acronym}
    \acro{AP}{access point}
    \acro{PTP}{Precision Time Protocol}
    \acro{OWD}{One-way delay}
    \acro{RTT}{Round-trip time}
    \acro{SMS}{Smart manufacturing systems}
    \acro{CPS}{Cyber-physical systems}
    \acro{IIoT}{Industrial Internet-of-Things}
    \acro{TUB}{Time Uncertainty Bound}
    \acro{UTC}{Coordinated Universal Time}
    \acro{GPS}{Global Positioning System}
    \acro{OWD}{One-way delay}
    \acro{ppm}{parts per million}
    \acro{RTT}{Round trip time}
    \acro{NIC}{Network interface controllers}
    \acro{DES}{Discrete event simulator}
    \acro{MAC}{media access control}
    \acro{TSN}{Time Sensitive Networking}
    \acro{NTP}{Network Time Protocol}
    \acro{COTS}{commercial off-the-shelf}
    \acro{PDV}{packet delay variation}
    \acro{LAN}{local area networks}
    \acro{SNMP}{Simple Network Management Protocol}
    \acro{SDN}{Software defined networking}
    \acro{NIC}{Network Interface Card}
    \acro{NC}{Network Calculus}
    \acro{MDR}{Maximum Drift Rate}
    \acro{TDD}{Time Division Duplex}
    \acro{NR}{New Radio}
    \acro{PHC}{PTP Hardware Clock}
    \acro{TT}{TimeTether}
    \acro{API}{Application Programming Interface}
    \acro{DNC}{Deterministic Network Calculus}
    \acro{SP}{strict-priority}
    \acro{IWRR}{Interleaved Weighted Round-Robin}
    \acro{RB}{resource blocks}
    \acro{UE}{User Equipment}
    \acro{OFDM}{Orthogonal Frequency Division Multiplexing}
    \acro{SRS}{Sounding Reference Signal}
    \acro{UL}{Uplink}
    \acro{RRC}{Radio Resource Control}
    \acro{RAN}{Radio Access Network}
    \acro{IoT}{Internet of Things}
    \acro{DT}{Digital Twin}
    \acro{O-RAN}{Open RAN}
    \acro{RIC}{RAN Intelligent Controller}
    \acro{KPM}{Key performance metrics}
    \acro{PaaS}{Platform as a service}
    \acro{DTDL}{Digital Twins Definition Language}
    \acro{SLA}{Service Level Agreements}
    \acro{ML}{Machine Learning}
    \acro{IE}{Information Elements}
    \acro{QoS}{Quality of Service}
\end{acronym}

\title{TwinRAN: Twinning the 5G RAN in Azure Cloud \\
\author{\IEEEauthorblockN{Yash Deshpande, Eni Sulkaj, Wolfgang Kellerer\\
\IEEEauthorblockA{ Chair of Communication Networks, Technical University of Munich, Germany \\
Email: \{yash.deshpande, eni.sulkaj, wolfgang.kellerer\}@tum.de}}
}
}
\copyrightstatement
\maketitle

\begin{abstract}
The proliferation of 5G technology necessitates advanced network management strategies to ensure optimal performance and reliability. 
\ac{DT}s have emerged as a promising paradigm for modeling and simulating complex systems like the 5G \ac{RAN}. 
In this paper, we present TwinRAN, a \ac{DT} of the 5G \ac{RAN} built leveraging the Azure \ac{DT} platform. 
TwinRAN is built on top of the Open RAN (O-RAN) architecture and is agnostic to the vendor of the underlying equipment.
We demonstrate three applications using TwinRAN and evaluate the required resources and their performance for a network with ~800 users and eight gNBs. 

We first evaluate the performance and limitations of the Azure DT platform, measuring the latency under different conditions. 
The results from this evaluation allow us to optimize TwinRAN for the DT platform it uses.
Then, we present the system's architectural design, emphasizing its components and interactions. 
We propose that two types of twin graphs be simultaneously maintained on the cloud. 
The first is for intercell operations, keeping a broad overview of all the cells in the network. 
The second twin graph where each cell is spawned in a separate Azure DT instance for more granular operation and monitoring of intracell tasks.   
We evaluate the performance and operating costs of TwinRAN for each of the three applications.
The TwinRAN DT in the cloud can keep track of its physical twin within a few hundred milliseconds, extending its utility to many 5G network management tasks - some of which are shown in this paper.   
The novel framework for building and maintaining a \ac{DT} of the 5G \ac{RAN} presented in this paper offers network operators enhanced capabilities, empowering efficient deployments and management.

\end{abstract}

\acresetall
\section{Introduction}
\label{sec:Introdcution}

The rapid adoption of new wireless communication technologies like 5G has significantly transformed the telecommunication landscape.
Central to its success is the \ac{RAN}, which provides the foundation for higher data rates, higher reliability, lower latency, and massive connectivity. 
Running various applications such as autonomous driving, smart cities, and \ac{IoT} on the same infrastructure is an impressive feature of the 5G technology.
This advancement comes at the cost of increased complexity in the 5G \ac{RAN}~\cite{RANComplexity}. 
Advanced interference management methods, handover mechanisms, and an increased number of cells are needed to meet the various requirements of emerging applications. 
One manifestation of the advanced complexity is the move towards \ac{O-RAN}, a vendor-independent \ac{RAN} architecture that allows for bespoke tailoring of the \ac{RAN} parameters for the location and applications\cite{intelligentORAN}. 
As we move towards 5G+ and promise even more ubiquitous connectivity, the complexity of \ac{RAN} is expected to increase even further~\cite{towards6gEra}. 
This complexity and dynamic nature of 5G \ac{RAN} poses a substantial challenge for network operators regarding deployment, management, and optimization~\cite{understandingORAN}.

\ac{DT}s, virtual replicas of physical systems, offer a promising approach to addressing these challenges~\cite{DTfor5G}. 
Operators can simulate, monitor, and optimize network performance in real-time by creating a \ac{DT} of the 5G RAN, enabling predictive management, efficient resource allocation, and rapid troubleshooting~\cite{corici_digital_2023, masarachiaDTORAN}. 
\ac{DT}s are being leveraged in many other industries, such as manufacturing, logistics, construction, and energy distribution~\cite{DTSurvey}. 
At the forefront of the \ac{DT} technology is the Microsoft Azure \ac{DT}s~\cite{nath2021building}. 
It is a \ac{PaaS} to create, maintain, and interact with \ac{DT}s on the cloud.
The Azure cloud, where the \ac{DT}s are hosted, is a production-grade cloud platform guaranteeing scalability, availability, and robustness. 
Moreover, other Azure cloud storage, analytics, and visualization services can be seamlessly connected to the \ac{DT}s. 
Therefore, Azure DT presents a compelling opportunity for twinning the 5G \ac{RAN}. 
Hence, we first measure the real-time performance of Azure DT under different conditions to make optimized design decisions for TwinRAN. 
The results presented from these measurements are a supplementary contribution to TwinRAN.  
To the best of our knowledge, this paper is the first to present a deep study of the behavior of Azure DT.

The \ac{O-RAN} standard is poised to play a crucial role in the future of 5G communication as it provides vendor diversity, cost-effectiveness, and efficiency~\cite{towards6GNetworks}. 
The \ac{O-RAN} standard already provides a method to monitor the network performance, user activity, and overall system health of the 5G \ac{RAN}~\cite{understandingORAN}. 
These messages are processed by the Near-Real-Time RAN Intelligent Controller (Near-RT RIC), which serves as a central hub for collecting data and controlling the 5G RAN.
To effectively utilize this data, we have developed a purpose-built xApp that interfaces with the Near-RT RIC. 
This xApp extracts and filters relevant information from the O-RAN indication messages and updates the \ac{DT} instance.  
This ensures that the virtual representation of the 5G RAN accurately reflects the current state of the physical network. 

The main \textbf{contribution} of this work is that it proposes a novel method and architecture for creating and managing a DT of the 5G RAN on the Azure cloud platform. 
Our approach leverages Azure DT's standardized modeling, data integration, and ORAN's non-invasive and vendor-agnostic properties. 
TwinRAN has three main features that underscore its practical utility: \textbf{Non-Invasive}: It does not require changes in the source code or hardware in the 5G RAN. By plugging the xApp on top of 5G RAN architectures, it builds seamlessly over the existing 5G components. 
\textbf{Vendor Agnostic}: It can work with all equipment and software vendors as long as they are \ac{O-RAN} compliant. 
\textbf{Multipurpose}: The same \ac{DT} can be simultaneously used for different applications as long as all the parameters required by the applications are modeled in the twins. Any application can connect to the \ac{DT} to stream real-time updates or load historical data. 
Furthermore, this property extends TwinRANs' utility for the future: any new application can be deployed without modifying the TwinRAN code.  
To present TwinRAN, it was first required to determine if the Azure DT platform suits this task.
Secondly, for TwinRAN to be efficient, it was also necessary to characterize the behavior of Azure DT (especially in terms of latency) under different packet sizes and twin sizes.  
Thus, results from a thorough measurement study to ascertain the latency characteristics of Azure DT are a \textbf{secondary contribution} of our work. 
To the best of our knowledge, this paper is the first to present a deep study of the behavior of Azure DT.

The rest of the paper is organized as follows: the next section provides the background on Azure DTs and the O-RAN system. 
Following the background, we mention some related work and differentiate our contributions.
In section \ref{sec:evaluating_adt}, we evaluate the behavior of Azure \ac{DT}. 
The learnings from this section are then used to present the architecture of TwinRAN in Section \ref{sec:architecture}.
This section also elaborates on the \ac{DTDL} models and the xApp design. 
Section \ref{Sec:Impl} addresses the implementation of TwinRAN and introduces the three use cases, and Section \ref{sec:results} evaluates the performance and required resources for TwinRAN, mainly focusing on pricing. 
Finally, in Section \ref{sec:conclusion}, we conclude, presenting the limitations and point toward out future work.

\section{Background}
\label{sec:background}

\subsection{O-RAN Standard}
\label{subsec:oran_standard}
The \ac{O-RAN} standard represents a significant evolution in the architecture of \ac{RAN}, aimed at fostering interoperability, flexibility, and innovation within the 5G ecosystem. 
The \ac{O-RAN} Alliance, a global consortium of network operators and vendors, spearheads this initiative to standardize open and intelligent RAN interfaces, enabling multi-vendor deployments and advanced network management capabilities~\cite{ORANArchi}.
The components and message structures of \ac{O-RAN} relevant to this paper are mentioned below. 
\subsubsection{E2 Node}
The E2 node intermediates the gNB and the Near-RT RIC~\cite{E2Oran}. 
It facilitates the exchange of telemetry data from the gNB to the Near-RT RIC and control messages in the opposite direction.  
This messaging enables the Near-RT RIC to manage and optimize RAN functions in real-time. 
TwinRAN uses the telemetry information received at the Near-RT RIC to update the DT. 

\subsubsection{Near-RT RIC}
The Near-RT RIC is designed to perform near-real-time control and monitoring of RAN elements~\cite{ORAN2021RIC}. 
It processes data collected from the E2 nodes and delivers it to the xApps. 
On the other hand, it also allows xApps to control the RAN elements over the E2 interface.  
Near-RT RIC usually contains a database that collects persistent information from various E2 agents. 

\subsubsection{xApps}
xApps are modular applications that run on the Near-RT RIC, providing targeted control and optimization capabilities. 
They are designed so that the vendors of the Near-RT RIC and the xApps can be disaggregated. 
These applications leverage the Near-RT RIC's real-time data and analytics capabilities to implement specific use cases ~\cite{ColoRAN, bonati2022intelligent, connectionManagementxapp, trafficsteerringxapp}, improving the system's efficiency. 
Some exemplary use cases are traffic steering, interference management, and resource optimization.
DTs is a novel use-case where a new DT updating xApp is required. 
The TwinRAN xApp filters data available at the near-RT RIC, checks for changes in the entities, and updates the twins on the cloud. 

\subsubsection{Message Types}
O-RAN defines several message types for communication between its components.
\textbf{Indication}: Provides status updates and telemetry data from RAN elements to the Near-RT RIC. The two types of indication messages are \textit{report} and \textit{insert}.
While the first gives periodic updates on subscription, the other is triggered asynchronously to notify the xApp of a specific event at the \ac{RAN}. 
\textbf{Policy}: Convey policies from the Near-RT RIC to guide the behavior of RAN elements. 
    The passing of such a policy function is reported to the TwinRAN instance by the TwinRAN xApp. 
\textbf{Control}: Used for configuring and managing RAN elements. While TwinRAN does not use control messages, we show an example in Section \ref{Sec:Impl} how they can be used in conjunction with TwinRAN to perform a handover. 

\begin{figure}
    \centering
    \includegraphics[width=\columnwidth]{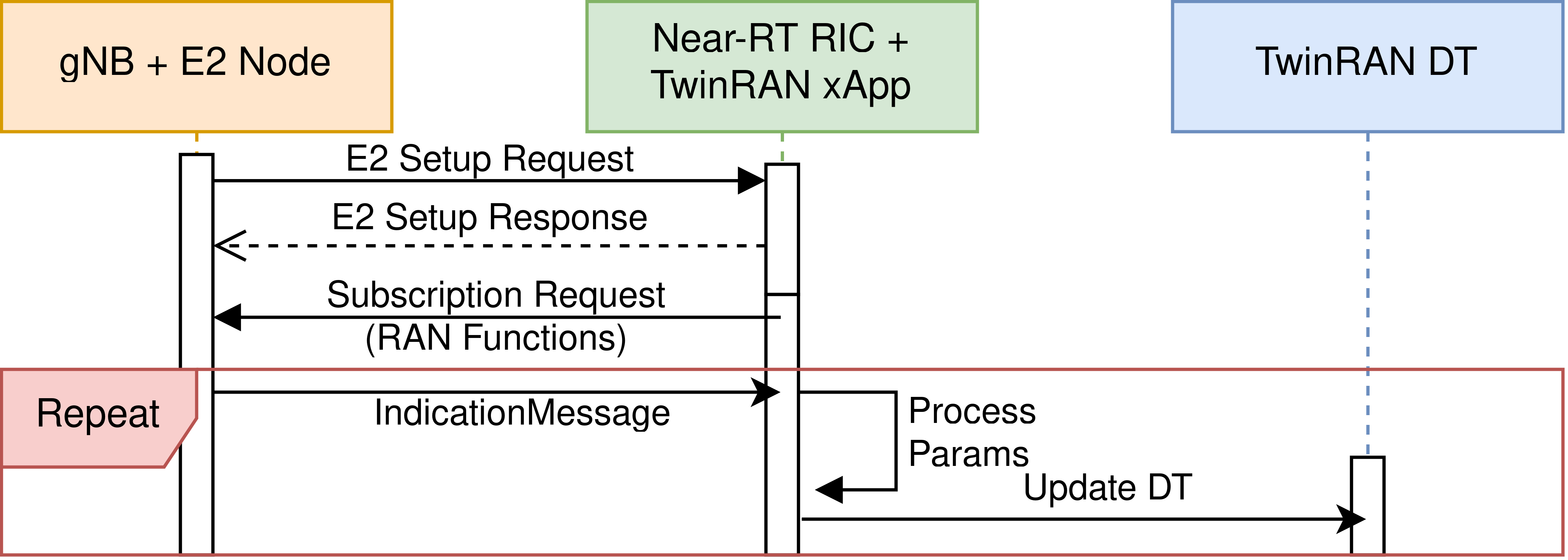}
    \caption{\small{Sequence Diagram of the messaging in TwinRAN: The TwinRAN xApp uses the O-RAN indication messages from the Near-RT RIC and updates the DT hosted on the Azure cloud. The desired KPMs have to be subscribed to by the xApp via the Near-RT RIC.}}
    \label{fig:sequencediagram}
\end{figure}
Figure \ref{fig:sequencediagram} illustrates how TwinRAN sits atop the \ac{O-RAN} architecture. 
The TwinRAN xApp uses the indication messages from the E2 node, then processes them to determine which messages need to be sent to the \ac{DT} instance on Azure cloud. 
The Near-RT RIC subscribes to the relevant \ac{RAN} functions available at the E2 Node. 
An important RAN function for TwinRAN is the \ac{KPM} defined in the \ac{O-RAN} standard~\cite{ORAN2021KPM}, giving us a good starting point on what can be represented using our proposed architecture. 

\subsection{Azure DTs and Cloud Services}
\label{subsec:background_azure}
Azure \ac{DT}s is a comprehensive \ac{PaaS} provided by Microsoft that enables the creation of digital representations of real-world entities and their relationships. 
This platform facilitates system modeling, monitoring, and optimization, leveraging other Azure cloud services seamlessly connected to it.  
Due to the standardized models and interfaces, the Azure DT is suited for large-scale systems~\cite{jacoby2020digital}. 

\subsubsection{Models}
Models in Azure DT define the structure and properties of DTs. 
These models are created using the \ac{DTDL}, specifying entities' characteristics, behavior, and capabilities within the system.
\ac{DTDL}, is JSON-like and comprises a set of metamodel classes ~\cite{AzureDigitalTwinsDTDL}. 
Defining clear and consistent models ensures a standardized representation of the physical entities and their attributes~\cite{meijers2022hands, jacoby2020digital}.
The main metamodel classes are: 
\textbf{Interface}: Defines the blueprint for a digital twin, outlining its structure, properties, and capabilities. 
\textbf{Property}: Describes attributes of a digital twin, holding specific static values. For example, the serial number of the thermometer.  
\textbf{Relationship}: Details the connections between digital twins, specifying how they interact and relate to each other—for example, the room and the thermometer and thermostat. 
\textbf{Component}: Represents a sub-entity of a digital twin by linking to an interface. It allows for modularity in defining models—for example, the window in a room. 
DTDL defines more classes, such as command and telemetry, which do not have a unique advantage in DTs~\cite{adttelemetry}
They are mainly useful for other Azure IoT services. 

\subsubsection{Twins and Twin Graph}
While models provide a static structure to defining \ac{DT}s, twins and twin graphs are instances of a model, or many models maintained in real-time on the cloud platform.
Twins in Azure \ac{DT} are the digital representations of physical entities. 
Each twin encapsulates all the defined properties and relationships of the physical system as defined in its model.  
These properties are maintained in an \textit{instance}, providing a virtual model that mirrors its real-world counterpart. 
This instance is hosted on a live execution environment. 
Twin graphs are the interconnected networks of twins that depict the relationships and interactions among different entities within a system. 
In Azure DT, twins are not isolated; they are linked through defined relationships, forming a comprehensive graph representing the entire system's topology and dependencies. 
This graph structure allows for modeling complex interactions and behaviors and keeping the system modular and scalable~\cite{knowledgegraphs, Akroyd_Mosbach_Bhave_Kraft_2021}. 

\subsubsection{Events and Notifications}
Events are occurrences or changes in the state of digital twins that trigger specific actions or workflows. 
Azure DT can generate and respond to events by connecting to the Event Hub, which processes and filters them.  
For example, we can store data in a database or update twins in another instance by routing and processing events. 
Events can also be directly processed to provide alerts and updates to the operators. 

\begin{figure}
    \centering
    \includegraphics[width=\columnwidth]{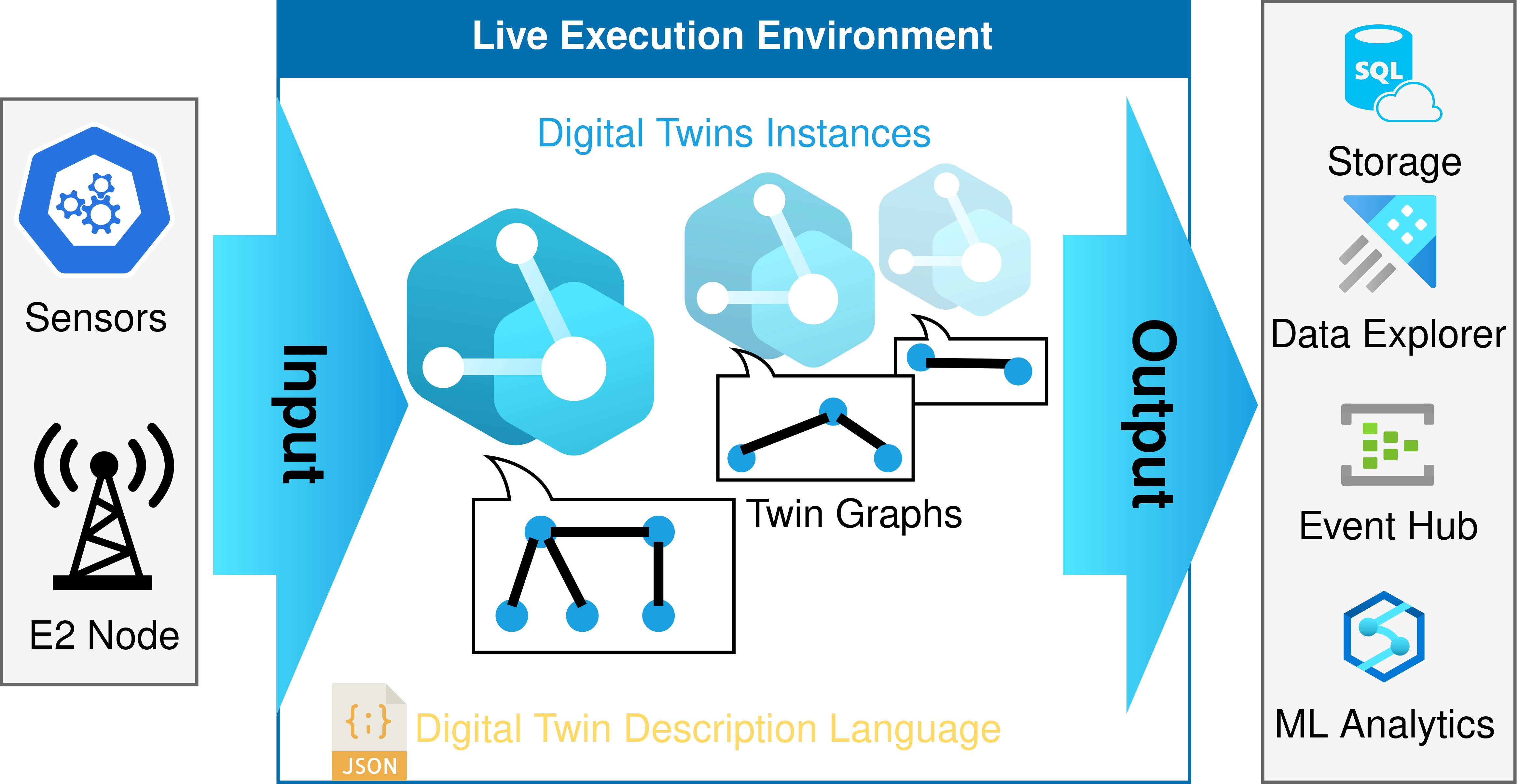}
    \caption{\small{Digital Twins on the Azure Cloud: In TwinRAN, the inputs to the DT instances hosted in a live execution environment are via the TwinRAN xApp. For every instance, the Twin graphs keep the twins and their relationship updated. The output of the DTs is then connected to other Azure cloud services for post-processing or recording. }}
    \label{fig:dt_architecure}
\end{figure}

Figure \ref{fig:dt_architecure} shows a general overview of TwinRAN on the Azure cloud. 
The E2 Node via the xApp and other related sensors push data to the DT instances.
The DT models, defined using DTDL, are used as a skeleton for each instance. 
These instances are executed in a live environment, and a twin graph maintains all of the twins' current statuses and dependencies. 
One can write an application to query the status of the DT instances in real-time or set up an automated stream of the DT changes to other Azure cloud services like Data Explorer or ML Analytics.

\section{Related Work}
\label{Sec:related_work}

\subsection{Digital Twins of Communication Networks}

Specialized DTs that facilitate the efficient operation of communication networks are presented in some previous works. 
For vehicular networks with 5G, ~\cite{digitalTwinv2x} provides an architectural framework but omits the \ac{RAN} models and focuses on the higher-level metrics such as throughput and latency. 
For simulating and optimizing load on edge computing servers using \ac{ML}, ~\cite{dtmec} uses a custom-built \ac{DT} of the CPU load on the server network. 
Similarly, ~\cite{vecnetworks} uses a \ac{DT} for simulating load on edge computing nodes on a vehicular network.
For an \ac{IoT} network using IEEE 802.15.4 protocol, a \ac{DT} using the Ecplise Ditto \cite{eclipseditto} is presented in ~\cite{DTN_Ecplisehono}. 
Indeed, this work provides a \ac{DT} similar to ours with vendor-agnostic, non-invasive, and multipurpose features. 
However, the 5G \ac{RAN} is much more complex than the \ac{IoT} network with low-powered devices mentioned in this work, and we tackle this complexity in this paper. 

For 5G network operation, \cite{mndtrodrigo, mahnoorcore} propose \ac{DT}s of the core network and the wired fronthaul network. 
In a service-based architecture such as the 5G core, where the network functions are packaged in docker containers, a generalized \ac{DT} for Kubernetes cluster optimization can also be used for such a purpose \cite{kapetanios, bhardwaj2022kubeklone}. 
We exclude twinning the core network from our current work and focus on the less-explored \ac{RAN}. 

A \ac{DT} to check if \ac{SLA}s are met is proposed in \cite{sun2021digital, sla_assurance}. 
Both works are mainly aimed at fault prediction and diagnosis to reduce network downtime. 
A \ac{DT} of the network for service resilience and security testing using chaos engineering technique \cite{chaoostwin}. ~\cite{dt_cyber_sec} focuses solely on the security of the 5G network by connecting a \ac{DT} for attack detection. 
A \ac{DT} for estimation and optimal configuration of data-center networks is presented in ~\cite{mimicnet, wisewhatif}.

Many works skip the explicit modeling of entities and learn their behavior using \ac{ML} techniques~\cite{mllrearningDT, mimicnet, RANDT_scenario, razvan}.
However, the learned \ac{DT} is not always accurate and requires a lot of data to be trained. 
While this approach is useful for twinning entities that are difficult to model or monitor in real-time, our approach in this work is to twin explicitly the entities in the \ac{RAN} that the Near-RT RIC can measure. 
As we show in this paper, using the \ac{O-RAN} architecture, we can have many properties of the 5G \ac{RAN} represented in their digital models, and the TwinRAN can be helpful in many different network operation tasks. 

In general, while specialized \ac{DT}s are useful, they are not efficient as operators need to deploy multiple custom-made \ac{DT}s in their network to fulfill different tasks. 
Hence, a multi-purpose twin, such as TwinRAN, reduces operation and maintenance costs compared to other state-of-the-art counterparts.
Hosting a \ac{DT} in a live execution environment in the Azure cloud and providing easily connectable end-points opens the possibility of seamlessly integrating additional applications in the future without changing the underlying twin models or implementation.  
Many works mentioned above also lack the respective \ac{DT}s implementation and operation details. 
We show that TwinRAN can be deployed in a production-grade cloud platform and discuss the operating costs and limitations in this paper.  
Finally, similar to this paper, the \textit{lag} of the twin of a factory floor in Azure \ac{DT} was measured in ~\cite{casepaper} as the API response times reported by the Azure DT dashboard. 
However, the end-to-end lag differs from the API response times, and we provide a more accurate measuring method for the lag. 
We also compare this lag for different twin and model sizes.
\subsection{Digital Twin Platforms}  
\textbf{Eclipse Ditto}:~\cite{eclipseditto}An open-source framework part of the Eclipse IoT Working Group. 
It provides the tools to create, manage, and interact with digital twins.
Similar to Azure DT, it has a JSON-based twin model defining language. 
However, twin instances are not maintained as knowledge graphs as in Azure \ac{DT}s. 
It can be self-hosted and connected to various MQTT and HTTP protocol endpoints.
\textbf{MCX}: ~\cite{shahsavari2021mcx} An open-source model description and updating method via MQTT protocol. 
It connects to simulation frameworks such as MATLAB, Ansys, or custom Python classes. 
It facilitates connecting simulation frameworks with real-time data.
\textbf{AWS IoT TwinMaker}:~\cite{awstwinmaker} AWS IoT TwinMaker is a cloud-native platform rivaling the Azure DT platform in its capabilities and features. 
It connects to other AWS services for monitoring, analytics, and data ingestion. 
The models are defined in a JSON-like language, and the twin instances are maintained in a knowledge graph. 
Many web tools like Grafana dashboards are available for twin visualization and monitoring. 

Due to its open-source nature, the Eclipse Ditto could be a good alternative for Azure DT. 
It could be more customizable and allow self-hosting within the 5G operators' infrastructure. 
However, it would also need more developer effort to connect the output of the \ac{DT} to other storage and analytics services that come out-of-the-box for Azure \ac{DT}. 
MCX could be useful for what-if scenario testing with other simulators such as MATLAB 5G Toolbox, next GSIM~\cite{albapaper} or ns-O-RAN~\cite{ns3_oran} but remains limited for real-time network operating tasks. 
We could not ascertain any considerable advantage for either Azure DT or AWS IoT TwinMaker in terms of features. 
They are also very similar in pricing~\cite{pricingcloud} and API response times~\cite{cloudcomparisons}.
A more positive user sentiment, due to its enterprise-wide integration with Windows and other Office services and its user-friendly dashboard, was reported for Azure \cite{sentimentscloud}.

\section{Evaluating Azure DT}
\label{sec:evaluating_adt}

\begin{figure}
    \centering
    \includegraphics[width=\columnwidth]{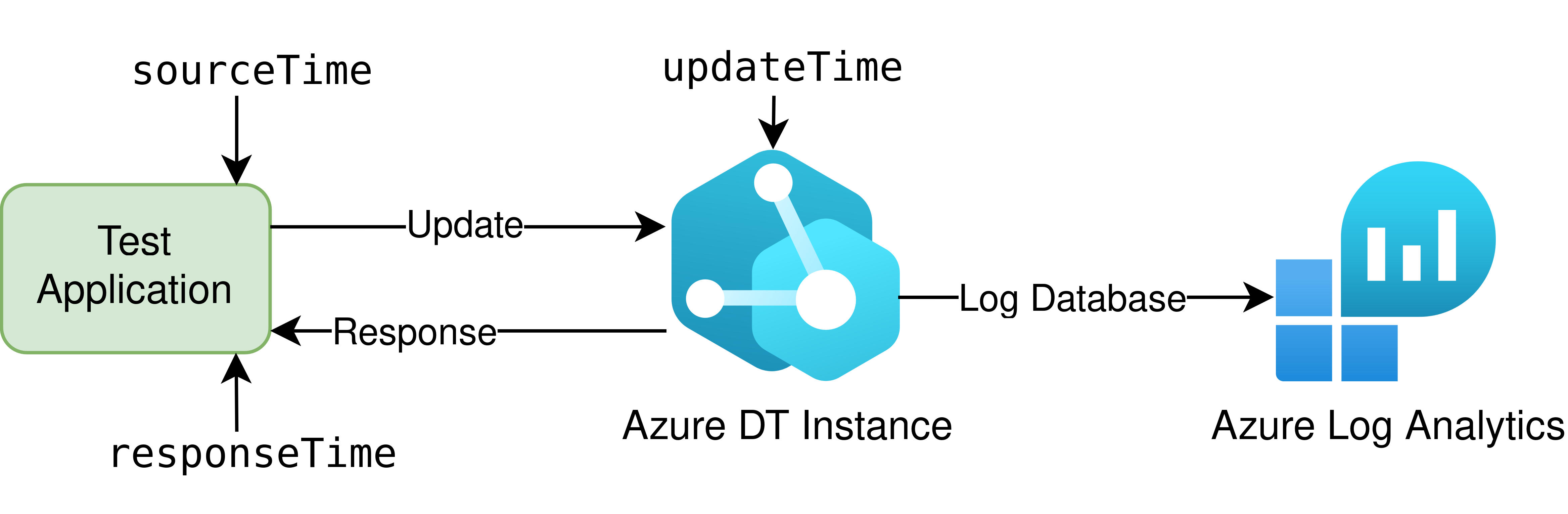}
    \caption{\small{Measurement setup for the \textit{lag} and query latency of Azure DT. The update messages are timestamped at three points. We vary model sizes and update sizes to ascertain the behavior of Azure DT.}}
    \label{fig:generic_mesurement_setup}
\end{figure}
\begin{figure*}
    \centering
    \includegraphics{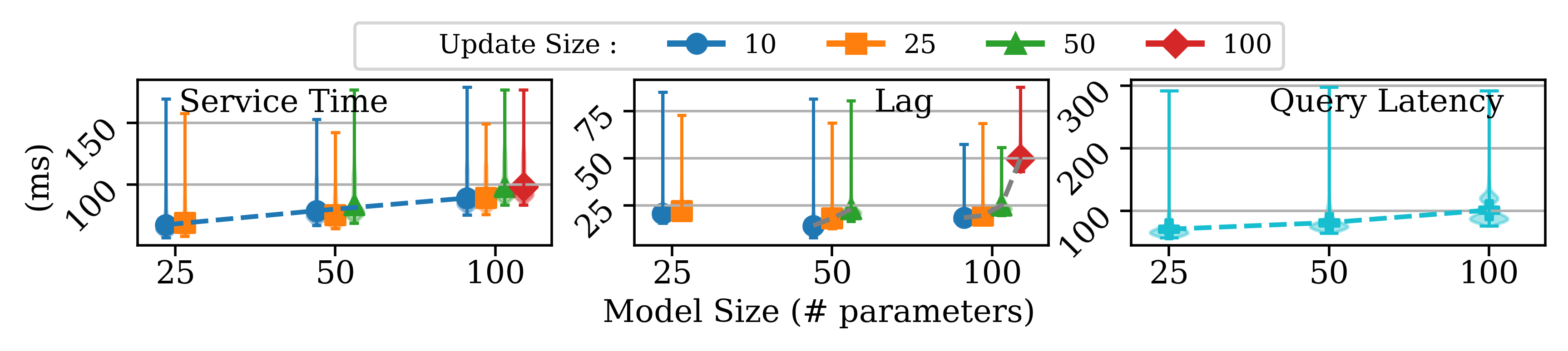}
    \vspace{-0.1cm}
    \caption{\small{Service times, lag, and query latency measurements for different DT models and update sizes. The service times scale with not only the size of the model but also the size of the update message. The measurements' mean values exhibit a linear trend, and the distributions have a long tail. }}
    \label{fig:evaluation_fig_1}
    \vspace{-0.5cm}
\end{figure*} 
In this Section, we asses the abilities of Azure DTs, mainly by looking at latency and lag of the \ac{DT} in the cloud for different model and update sizes. 
A deep investigation of Azure DTs in this regard is missing from the state of the art.
The primary goal of the \ac{DT} is to keep track of the changes in its physical twin. 
We define the \textit{lag} as the time it takes for this change to be reflected in the \ac{DT}. 
Figure \ref{fig:generic_mesurement_setup} shows the measurement setup for the evaluation. 
A simple Python application connected to the Azure DT instance creates twins, updates them, and generates queries. 
Additionally, all the changes in the Azure DT instance are streamed to an Azure Log Analytics dashboard, which provides timestamped logs of API calls and their internal latency. 
For every property in Azure DT, the \ac{DTDL} model has two inbuilt attributes: \code{lastUpdatedTime} and \code{sourceTime}. 
The \code{lastUpdatedTime} is when Azure DT updates the property, and the \code{sourceTime} is an optional value that can be appended to every property by the source. 
The \ac{DT} instance responds to the API client in the test application, indicating whether the update was successful and detailing the reasons for any failures.
We introduce another timestamp called the \code{responseTime} of the instant when an update API call is returned. 
Another API call to Azure DT is the query call. 
This provides the current state of the twin in Azure DTs. 
The test application had a separate program querying the models, ensuring they did not conflict with the update messages. 
The test application was hosted on a 32-core server with a 10G connection to the gateway, and the system time was synchronized by \ac{NTP}.
We increased the \ac{NTP} poll rate to the maximum (1/64th of a second), giving us sub-millisecond time synchronization accuracy between the cloud and the source.
We define the lag as the mean of \code{lastUpdatedTime} - \code{sourceTime} for all the parameters of a twin. 
While we define the service time of an update as the \code{responseTime} - \code{sourceTime}. 
We create three \ac{DTDL} models with 25, 50, and 100 parameters each. 
Then, we updated/queried each of the models 10,000 times with update sizes of 10, 25, 50, and 100 parameters to the models that can accommodate them. 
The updates and queries were generated in batches of 500 at different times of the day, and each update was sent 5 seconds apart. 

Figure \ref{fig:evaluation_fig_1} shows the service times, lag value, and query times for different models and update sizes.
The mean values are highlighted on the violin plot. 
The mean service times increase linearly with model size \textit{and} update size. The same size of an update to a larger model requires longer to serve.
The increase in service times is roughly 2.8ms/10params, while the smallest value of service time was observed to be 45ms, and no update took longer than 200ms to service. 
While a bigger update exhibits a more significant lag, there is no discernable difference in the lag value for different models with the same update size. A large or small model takes the same amount of time on average to update the same number of parameters.
The mean lag increases linearly at roughly 2.8ms/10params. The shortest lag time was 9ms, while at no time was any twin lagging with more than 100ms.
Querying larger models takes longer. The mean query latency increases roughly at the rate of 4ms/10params. The smallest querying time was 60ms. 
We also observe that the values' distribution exhibits a long tail. This is perhaps due to such behavior of object storage systems~\cite{cloudlatency}. While the mean values show a specific trend depending on the update size, the maximum values of both lag and service times are not dependent on the model or update size.

To ascertain whether any two distributions in our measurements are the same, we employ the Kolmogorov-Smirnov test with a p-value threshold of 0.05. 
The time of the day did not make any difference to any of the three values mentioned above.
Similarly, updating or querying multiple twins on the same Azure DT instance did not increase either of the three values. 
When the model and update sizes were 100, we measured no difference between the first and last parameters' lag. 
This indicates that the parameters on a twin are either not updated sequentially but rather parallelly, or the time to update all the parameters is very small compared to the overall lag and service time. 
The performance in terms of lag was highly scalable in terms of the number of twins. 
We did not notice any difference in the lag when whether there was one twin or 100 twins/instance. 
This is expected as the event route endpoint of each twin is different internally. 
Hence there is no pipeline blocking a twins update when parallelly updating a large number of twins at the same time.
For $Y$ twins to be updated, it is like sending $Y$ different API calls. 

We perform another test, sending two successive updates or queries every 5 seconds.
The second message is sent 5 ms after the first. 
It was observed that the second update's mean lag, service time, and query time were 45 ms more than the first. 
This result indicates that Azure DT locks access to a twin when a read or write operation is performed. 

As mentioned in Section~\ref{subsec:background_azure}, \ac{DTDL} also allows us to define \textit{relationships} indicating how two twins connect.
Like a twin update, any relationship can be updated with a separate API call to the Azure DT instance. 
A relationship update, on average, takes 10ms more than a twin update of the same size. 
Finally, it was observed that the time to create a new twin on the Azure DT instance from a given model also increases linearly with the model size.  
However, for the 3 model sizes mentioned in this section, this value was never smaller than 1 second and larger than 2 seconds. 

Azure DT provides certain service limits to its users\cite{adtservicelimits}. 
The maximum size of the JSON string in the update message with 100 parameters was 6300 bytes. 
The service limit for this value is 32KB, which provides enough headroom. 
Azure DT's other relevant service limit only allows 10 updates to any twin per second. 
At the same time, one Azure DT instance can handle 1000 updates per second. 
Thus, having larger models and larger updates would be beneficial as we can pack more information in the DT without violating the service limits. 
This comes at the cost of added latency.
In the next section, we present our approach in TwinRAN to balance these two goals. 

\section{TwinRAN Models and Architecture}
\label{sec:architecture}

\subsection{Intercell and Intracell Instances}
\label{subsec:model_sizes_etc}
The limits placed by Azure DT provide us with considerable headroom on the message size.
Let the maximum number of twins in an instance be $Y \in \mathbb{N}^{+}$.
If every twin in the instance is updated once every $R \in \mathbb{R}^{+}$ seconds, then the Azure DT constraint is $R \geq 0.1$, and we get $Y=1000\cdot R$. 
5G RAN systems are dynamic, and the maximum frame and slot length are 10 and 1 ms, respectively.  
Azure DT was initially built to represent the physical world using IoT, where the changes in the environment are slower than 5G. 
To maintain the fidelity of the DT in the cloud in TwinRAN, the update rate must be as high as possible within the limits placed by Azure DT.
Updating all the twins at the maximum rate of $R=0.1$ seconds, we can have a maximum of 100 twins per instance.
Any \textit{relationship} can be defined between two twins in the same instance.
Thus, a problem arises in organizing the twins to achieve scalability while keeping to 100 twins per instance limitation and keeping all the relationships between entities.
At the same time, the models' size must be kept small to decrease the lag of the twins.  
The models must also be organized to fit the entities traditionally used in cellular networks. 
Hence, we propose to split TwinRAN into two instance types. 

From the perspective of a 5G RAN operator, the tasks can be divided into two main categories: intracell and intercell. 
Intracell operation tasks include efficient scheduling of users based on their required \ac{QoS}, minimizing operating costs, network security, etc. 
Intercell operations are handover management, interference mitigation, etc.  
Therefore, we propose two twin-graphs simultaneously maintained in the cloud: TwinRAN-cell and TwinRAN-multi. 
TwinRAN-multi keeps a broad overview of the entire network. 
Each cell is defined in one \ac{DTDL} model and neighboring cells are connected with \textit{relationships}.
Each UE is one property of the gNB, with SINR as its value. 
TwinRAN-cell spawns a new instance for every gNB. 
Here, the gNB and UE are two different \ac{DTDL} models with a relationship between them if the UE is connected. 
We present the DTDL models in detail in the following subsection.
A large model of a cell with detailed models of the gNB and UE as \textit{components} would allow us to simultaneously manage 100 cells, each updating 10 times a second.   
However, a cell containing 100s of UEs would make the update sizes large and increase the twins' lag, service time, and query time. 
Moreover, this would impact the system's scalability as an additional parameter in the UE model and severely affect the overall lag. 
We can keep these values low by employing a disaggregated approach that simultaneously maintains TwinRAN-cell and TwinRAN-multi instances. 
\begin{figure}
    \centering
    \begin{subfigure}[t]{0.4\columnwidth}
        \centering
        \includegraphics[trim={0 0 0 26cm},clip,width=0.8\columnwidth]{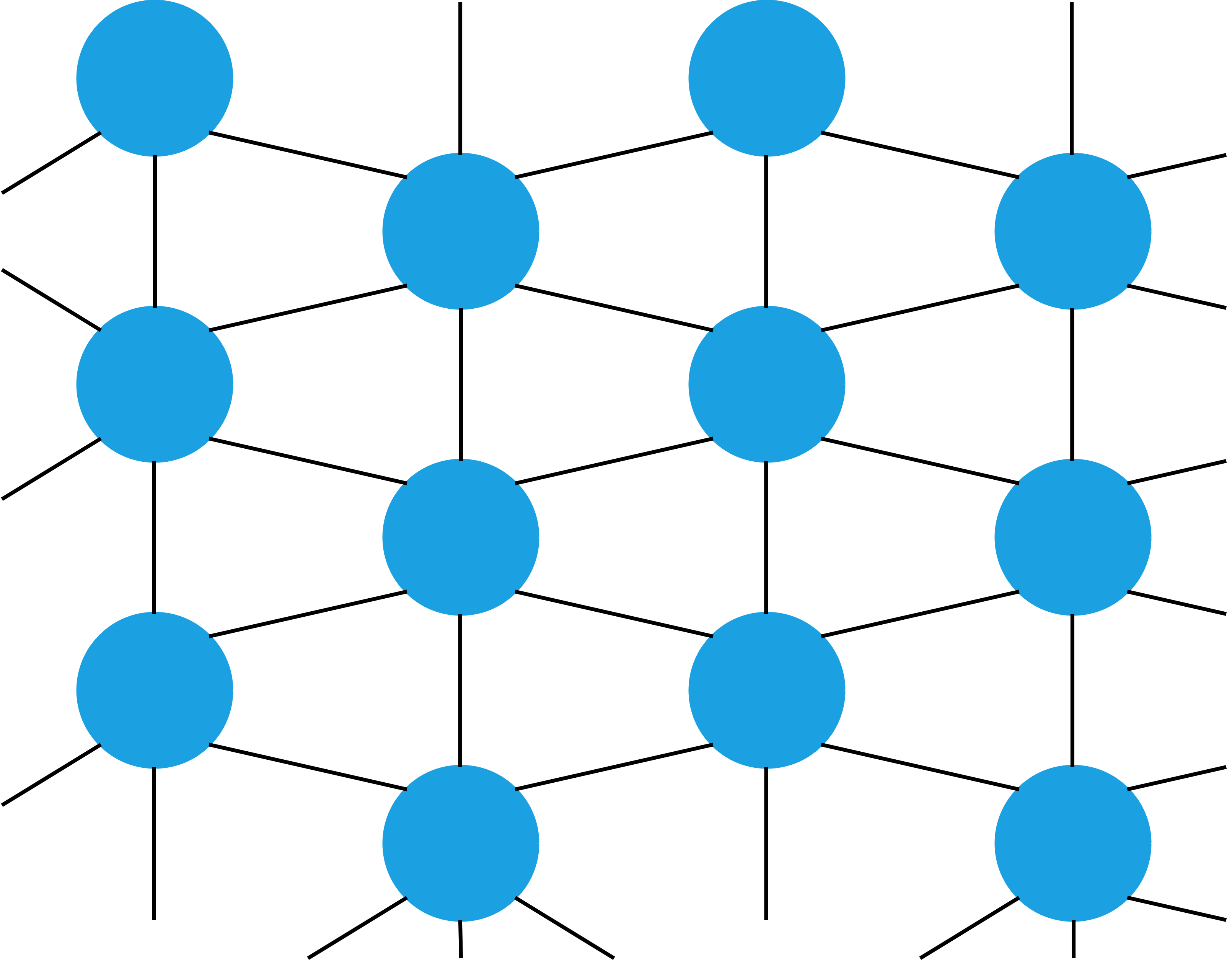}
        \caption{}
    \end{subfigure}%
    ~ 
    \begin{subfigure}[t]{0.6\columnwidth}
        \centering
        \includegraphics[width=0.8\columnwidth]{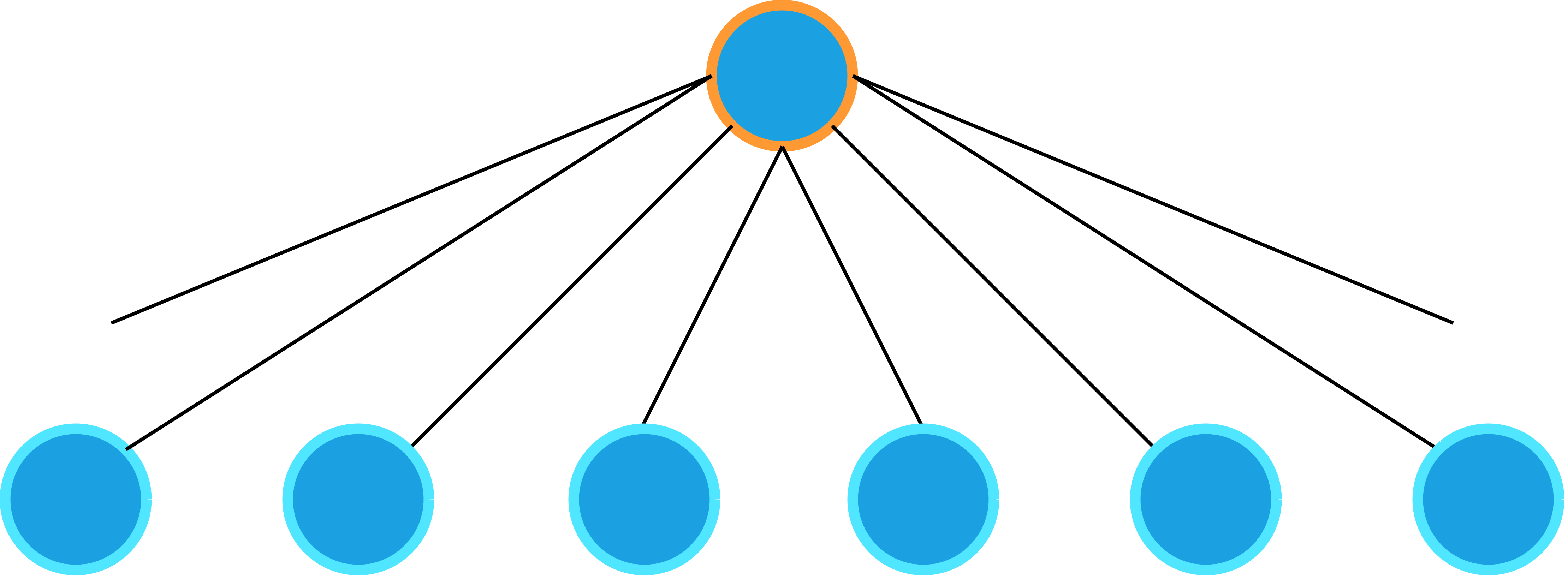}
        \caption{}
    \end{subfigure}
    \caption{\small{Two different twin-graphs are maintained : (a)TwinRAN-multi: a single instance keeps track of the entire network. One DTDL model defines one cell. The neighboring cells are connected by relationships, mainly defining the amount of interference caused by the neighbor. (b) TwinRAN-cell, where one instance is spawned for every cell. The gNB connects to each UE via a relationship.}}
\end{figure}

\subsection{DTDL Models}
As mentioned in Section~\ref{subsec:background_azure}, models defined by DTDL provide a structure to each twin. 
TwinRAN defines three models: the cell, \ac{UE}, and the gNB. 
The cell is used in TwinRAN-multi. 
The input to our modeling step is  the information we can obtain via different O-RAN messages mentioned in Section~\ref{subsec:oran_standard}. 
The information is divided into \ac{IE}s, which are structured data units used within the \ac{O-RAN} protocol messages to convey specific parameters.
While the amount of information is substantial, and thus, one can make very detailed models, we make the smallest possible model useful for all the three use cases mentioned in Section \ref{Sec:Impl}.

Table \ref{tab:dtdl_models} shows the two DTDL \textit{interfaces} for the TwinRAN-multi instance. 
The gNB is already initialized with the maximum number of UEs it can accommodate (99),  regardless of their connection. 
This is because changing a model in the live execution environment requires more than 1 second, and hence, we do not need to change the model for every handover.
Neighboring gNBs are connected with a \textit{relationship} whose value is the received power from that gNB. 
The properties of the UE that are important to demonstrate our use cases are its location and received power. 

Table \ref{tab:dtdl_models} shows the two DTDL \textit{interfaces} for the TwinRAN-cell instance in colored. 
Here, a more detailed model is created for the UE.  
In DTDL, the datatypes of each property are called \textit{schema} and must be defined beforehand. 
Optionally, we can also define a range for each attribute. 
When the Azure DT API updates the twin in the live execution environment, the type and value check is done every time.
An error is returned to the update API if they do not match the ones in the DTDL interfaces. 
This enables the Azure DT to keep error-free and consistent models and ensures that a rogue update is not reflected onto the DTs.
\begin{table}[]
\resizebox{\columnwidth}{!}{%
\begin{tabular}{ccc|ccc}
\hline
\multicolumn{3}{|c|}{\textbf{gNB}} &
  \multicolumn{3}{c|}{\textbf{UE}} \\ \hline
\multicolumn{1}{|c|}{\textbf{Property}} &
  \multicolumn{1}{c|}{\textbf{Schema}} &
  \textbf{\begin{tabular}[c]{@{}c@{}}Range or\\ Candidate\end{tabular}} &
  \multicolumn{1}{c|}{\textbf{Property}} &
  \multicolumn{1}{c|}{\textbf{Schema}} &
  \multicolumn{1}{c|}{\textbf{Range}} \\ \hline
\multicolumn{1}{|c|}{PLMN} &
  \multicolumn{1}{c|}{int} &
  20-50 &
  \multicolumn{1}{c|}{RNTI} &
  \multicolumn{1}{c|}{int} &
  \multicolumn{1}{c|}{0-999} \\ \hline
\multicolumn{1}{|c|}{ARFCN} &
  \multicolumn{1}{c|}{int} &
  600000 – 2016666 &
  \multicolumn{1}{c|}{Location} &
  \multicolumn{1}{c|}{Geospatial} &
  \multicolumn{1}{c|}{-} \\ \hline
\multicolumn{1}{|c|}{Relationship} &
  \multicolumn{1}{c|}{float} &
  gNB &
  \multicolumn{1}{c|}{RSRP} &
  \multicolumn{1}{c|}{int} &
  \multicolumn{1}{c|}{30.0 - 160.0} \\ \hline
\multicolumn{1}{|c|}{UE1} &
  \multicolumn{1}{c|}{Component} &
  UE &
  ------------- &
  --------------- &
  ------------------ \\ \hline
\multicolumn{1}{|c|}{UE2} &
  \multicolumn{1}{c|}{Component} &
  UE &
  \multicolumn{1}{c|}{\cellcolor[HTML]{EFEFEF}Buffer} &
  \multicolumn{1}{c|}{\cellcolor[HTML]{EFEFEF}int} &
  \multicolumn{1}{c|}{\cellcolor[HTML]{EFEFEF}0-255} \\ \hline
\multicolumn{1}{|c|}{...} &
  \multicolumn{1}{c|}{...} &
  ... &
  \multicolumn{1}{c|}{\cellcolor[HTML]{EFEFEF}UL BLER} &
  \multicolumn{1}{c|}{\cellcolor[HTML]{EFEFEF}float} &
  \multicolumn{1}{c|}{\cellcolor[HTML]{EFEFEF}0.0-1.0} \\ \hline
\multicolumn{1}{|c|}{UE 99} &
  \multicolumn{1}{c|}{Component} &
  UE &
  \multicolumn{1}{c|}{\cellcolor[HTML]{EFEFEF}UL CQI} &
  \multicolumn{1}{c|}{\cellcolor[HTML]{EFEFEF}int} &
  \multicolumn{1}{c|}{\cellcolor[HTML]{EFEFEF}0-15} \\ \hline
\multicolumn{1}{l}{----------------------} &
  \multicolumn{1}{l}{-----------------} &
  \multicolumn{1}{l|}{---------------------------} &
  \multicolumn{1}{c|}{\cellcolor[HTML]{EFEFEF}DL BLER} &
  \multicolumn{1}{c|}{\cellcolor[HTML]{EFEFEF}float} &
  \multicolumn{1}{c|}{\cellcolor[HTML]{EFEFEF}0.0-1.0} \\ \hline
\rowcolor[HTML]{EFEFEF} 
\multicolumn{1}{|c|}{\cellcolor[HTML]{EFEFEF}Tx Power} &
  \multicolumn{1}{c|}{\cellcolor[HTML]{EFEFEF}float} &
  20-50 &
  \multicolumn{1}{c|}{\cellcolor[HTML]{EFEFEF}DL CQI} &
  \multicolumn{1}{c|}{\cellcolor[HTML]{EFEFEF}int} &
  \multicolumn{1}{c|}{\cellcolor[HTML]{EFEFEF}0-15} \\ \hline
\rowcolor[HTML]{EFEFEF} 
\multicolumn{1}{|c|}{\cellcolor[HTML]{EFEFEF}Relationship} &
  \multicolumn{1}{c|}{\cellcolor[HTML]{EFEFEF}int} &
  UE &
  \multicolumn{1}{c|}{\cellcolor[HTML]{EFEFEF}UL MCS} &
  \multicolumn{1}{c|}{\cellcolor[HTML]{EFEFEF}int} &
  \multicolumn{1}{c|}{\cellcolor[HTML]{EFEFEF}0-28} \\ \hline
\rowcolor[HTML]{EFEFEF} 
\multicolumn{1}{|c|}{\cellcolor[HTML]{EFEFEF}Connected UEs} &
  \multicolumn{1}{c|}{\cellcolor[HTML]{EFEFEF}int} &
  0-99 &
  \multicolumn{1}{c|}{\cellcolor[HTML]{EFEFEF}UL MCS} &
  \multicolumn{1}{c|}{\cellcolor[HTML]{EFEFEF}int} &
  \multicolumn{1}{c|}{\cellcolor[HTML]{EFEFEF}0-28} \\ \hline
\end{tabular}%
}
\caption{\small{DTDL \textit{Interfaces} for the TwinRAN-multi \& TwinRAN-cell (colored) instances. TwinRAN-multi: The UE is contained in the gNB using the DTDL \textit{component} metaclass, and all the parameters here are contained in a single twin. TwinRAN-cell: Every instance has one gNB and $N\leq 99$ UE twins.}}
\label{tab:dtdl_models}
\end{table}
\vspace{-0.1cm}
\subsection{TwinRAN xApp}
\label{subsec:xapp}
\begin{figure*}
    \centering
    \begin{subfigure}[t]{0.63\columnwidth}
        \centering
        \includegraphics[width=\columnwidth]{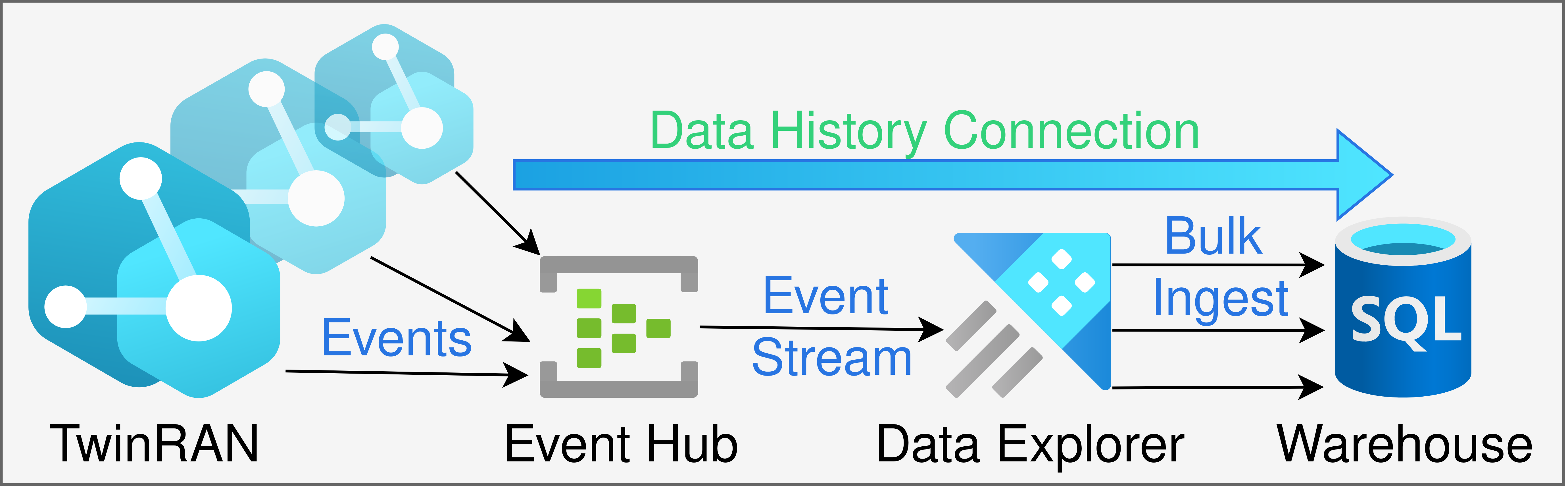}
        \caption{}
        \label{fig:use_case_1}
    \end{subfigure}%
    ~ 
    \begin{subfigure}[t]{0.63\columnwidth}
        \centering
        \includegraphics[width=\columnwidth]{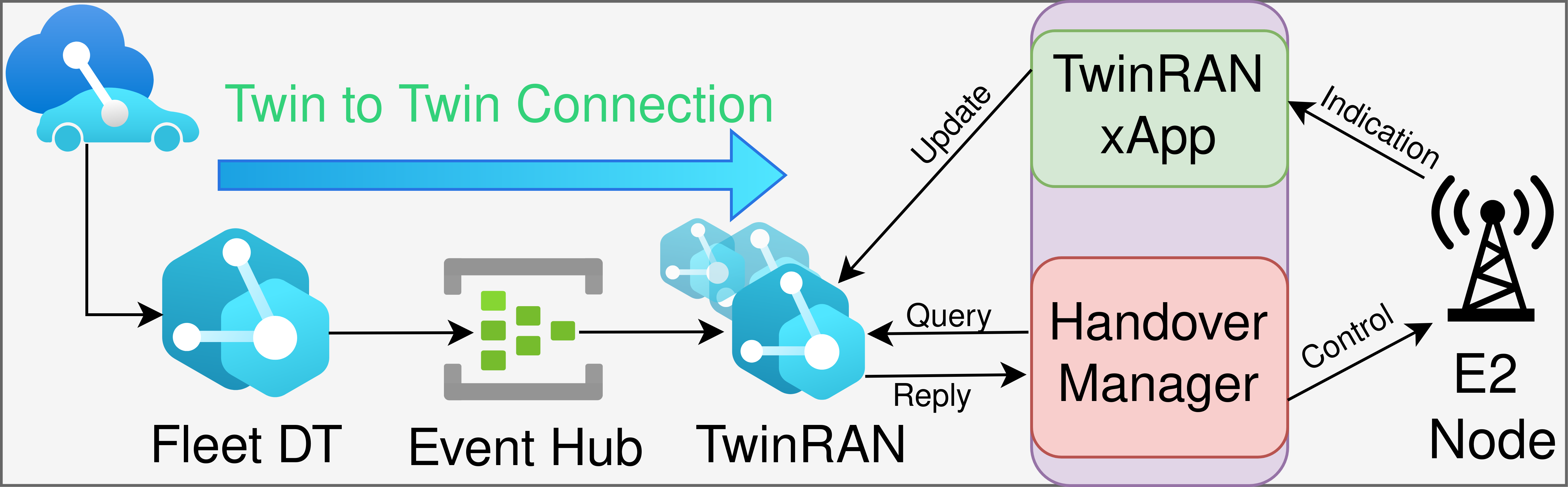}
        \caption{}
        \label{fig:use_case_2}
    \end{subfigure}
    ~ 
    \begin{subfigure}[t]{0.63\columnwidth}
        \centering
        \includegraphics[width=\columnwidth]{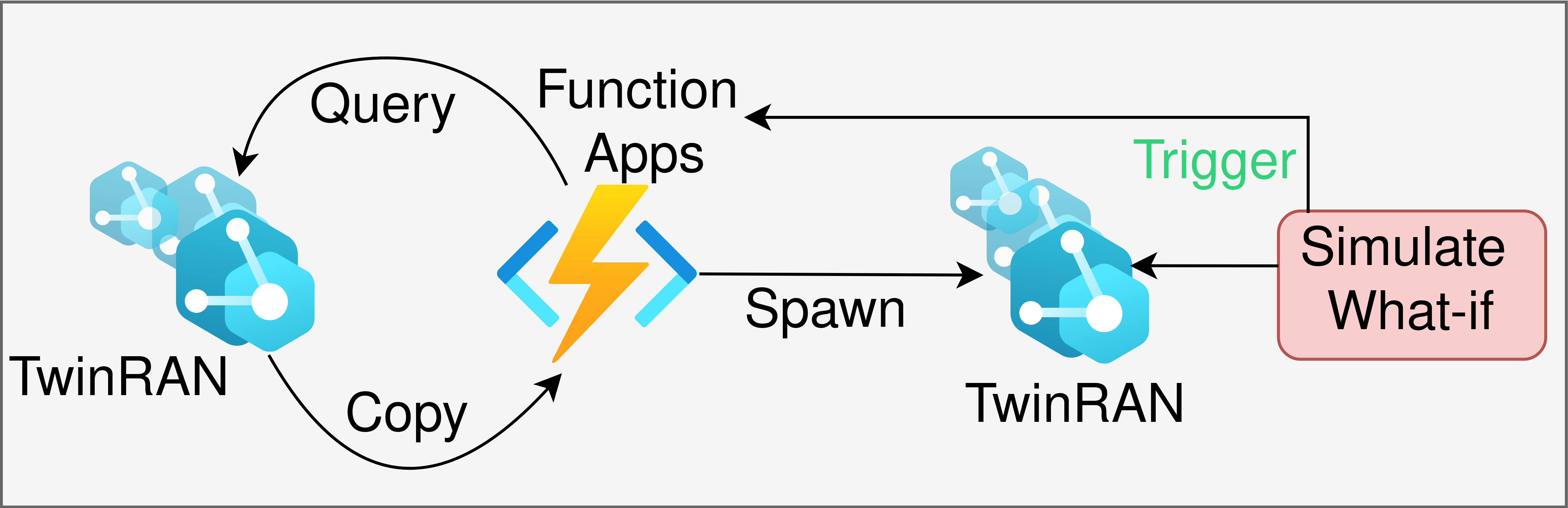}
        \caption{}
        \label{fig:use_case_3}
    \end{subfigure}
    \caption{\small{The use-cases with TwinRAN : (a) Data Warehouse for ML: a single instance keeps track of the entire network. One DTDL model defines one cell. The neighboring cells are connected by relationships, mainly defining the amount of interference caused by the neighbor. (b) TwinRAN-cell, where one instance is spawned for every cell. The gNB connects to each UE via a relationship.}}
    \vspace{-0.4cm}
\end{figure*}
The TwinRAN xApp updates the DT on the cloud and manages the instances, creating twins and maintaining relationships. 
The information obtained from the E2 Node is provided by different RAN functions~\cite{ORAN2021KPM}.
TwinRAN subscribes to the required RAN functions via the Near-RT RIC as shown in Figure~\ref{fig:sequencediagram}.
For periodic indication messages(called as \textit{report}, see Sec. ~\ref{subsec:oran_standard}), the subscription request also contains a parameter called \textit{granularity}.
This parameter represents the frequency at which the \textit{report} messages should be sent by the E2 Node. 
The gNB in the E2 setup request provides a cell ID and PLMN information used to spawn a new cell in the TwinRAN-muli instance.
Meanwhile, another instance is created in the Azure DT cloud, where the gNB twin is created. 
One expects this to happen rarely, as cells do not frequently appear and disappear from the operator's network. 

The xApp filters and packages relevant data to create API calls, which are then used to update the digital twin hosted on the Azure DT platform.
Although the xApp can only update the twin 10 times a second, the \textit{granularity} can be lower for the xApp to E2 node connections.
Thus, specific values can be averaged at the xApp before being updated at the twin. 
Apart from the RNTI, location, and number of connected UEs from Tables~\ref{tab:dtdl_models}, we send the average of the collected values since the last update - if this value has changed. 
The API call to the Azure DT live execution environment is made by packaging the update for each twin into a JSON string. 
The TwinRAN xApp synchronizes its clock with universal time via NTP and appends the \code{sourceTime} of every updated property.
In case of handovers, the TwinRAN xApp must delete a relationship and establish a new one with a different gNB twin or a different Azure DT instance.
At the same time, it must delete the same UE from the properties of the TwinRAN-multi instance and add it to the appropriate cell. 

\section{Implementation}
\label{Sec:Impl} 
\subsection{Use Case I: Data Warehouse for Machine Learning}
\label{subsec:usecase1}
A data warehouse is a centralized repository that stores large volumes of structured data from multiple sources.
Access to such vast amounts of structured data allows for conceptualizing training \ac{ML} models. 
PEACH~\cite{serkut_peach} predicts channel quality based on images. 
If one has the CQI data for many UEs from TwinRAN and timestamped footage from a monitoring camera, one could train the PEACH model for their setup. 
Furthermore, one can adopt a continuous training approach by periodically training on a new dataset. 
Figure \ref{fig:use_case_1} shows the architecture in the cloud for deploying the use-case. 
The Azure DT instances produce \textit{events} whenever a twin or a relationship is updated or created. 
These events are piped to an Azure Event Hub instance, a data streaming, processing, and ingestion platform. 
The processed events are streamed to the Azure Data Explorer instance, a data storage and analytics engine with an inbuilt persistent database. 
A Kusto Query Language (KQL) query from any platform can load data from the database to an ML model.  

\subsection{Use Case II: Precise Positioning for Handovers}
\label{subsec:usecase2}
Precise positioning can help the network make better handover decisions. 
However, the triangulation accuracy based purely on the NR signals is much less compared to GPS or LiDAR~\cite{v2xpositioning, positioingindoor}. 
Some recent works have proposed efficient handover management using O-RAN \cite{prado2024enhanced, traffic_steering}.
TwinRAN can provide a streamlined way to provide input to these algorithms. 
As DT deployment becomes more ubiquitous, a cross-connection between different DTs is inevitable. 
In this use case, we imagine an operator of a vehicle fleet or AGVs deploying their own DT. 
This comprehensive DT is very large and contains information like battery level, load carried, and distance run, along with the vehicle's current location. 
The network operator subscribes to a particular data stream (location coordinates) relevant to its operation and does not need to maintain their own position collection mechanism. 
Figure \ref{fig:use_case_2} illustrates the architecture to realize the use case.
Azure DT allows you to create \text{event routes} that connect their input and output (called as \textit{endpoints}) to the event hub.
The publisher of events can publish their data to the event hub with a unique ID that another event route can connect its endpoint to. 
The xApp that triggers the handover queries the channel quality, buffer status, interference information, and the location from the TwinRAN DT to feed it to its algorithm, oblivious to the source of the location data.  

\subsection{Use Case III: What-if Scenarios}
\label{subsec:usecase3}
The DT in the cloud is the closest representation of the current digital state of the system.
Network robustness and security can be greatly improved by conducting a what-if analysis by simulating different scenarios using a previous state of the network~\cite{wisewhatif, what-ifbgp,ak2024if}.
Figure \ref{fig:use_case_3} depicts the architecture for such a system using TwinRAN. 
A copy of the current TwinRAN state is made by simultaneously querying the entire instance and deploying it on many new instances. 
Azure Functions is a serverless computing service triggered by an event or an API call to make such one-to-many copies. 
The new instances of TwinRAN can then be connected to a simulation to conduct different what-if scenarios and compare the evolved twins.  
The vendor-agnostic capability of TwinRAN allows for such a seamless connection. 
We first populate a TwinRAN instance by the Flexric\cite{Flexric} with one real OpenAirInterface\cite{openairinterface} gNB and 7 E2 agent emulators.
The newly spawned copies of the TwinRAN instance were then connected to an NS-3 \cite{ns3_oran} simulation to test what-if scenarios 
\footnote{Some changes had to be made at the RIC-TwinRAN xApp interface depending on the three tested O-RAN RICs - Flexric, NS3-ORAN, and Near-RT RIC from the O-RAN software community. This is a limitation because true interoperability is still not achieved by any ORAN implementation or vendor and until this original goal of ORAN is achieved, small vendor-specific changes will need to be made at all points in the data chain. }.

\section{Results}
\label{sec:results}
We build a RAN network with $M=8$ cells with $N=99$ UEs in each cell. 
Thus, we have eight twins in the TwinRAN-multi instance and 100 twins in each of the 8 TwinRAN-cell instances. 
Each cell in the TwinRAN-multi instance is updated five times a second, while each Twin in the TwinRAN-cell instance is updated 10 times a second.
For each of the three use cases, we define relevant KPIs that we measure. 
For the first use case, the volume of data that traverses a data history connection per hour and is stored in the database is measured.   
The lag from the \code{sourceTime} to the TwinRAN \code{updateTime} is measured for use case II. 
Finally, the time it takes to spawn five new copies of the TwinRAN instance is measured for the last use case. 
We also present a cost analysis taking the pricing information by Azure for the Eastern United States for all three applications. 
The Azure DT generates 30 million messages and 4 million operations per hour in operation mode. 
As of the writing of this paper, Azure charges 1\$ per million messages and 2.5\$ per million operations, giving the base running price of our network as \textbf{40\$ per hour}. 
 
For the first use case, the xApp generates 240kB of data, and the event hub generates 64,000 events per second on average for the first use case.  
This was measured by monitoring the output port of the xApp and the logs from Azure event hub. 
The data explorer caches each event of ~21 bytes and batch-ingested into the database once every 5 minutes. 
Thus, approx. 4MB of data is ingested every 5 minutes. 
For such small amounts of data, a single event hub instance and 8vCPUs on the data explorer cluster are sufficient, as the CPU and cache utilization never exceeded 20 percent throughout our experiment.
The storage space of 1.5 TB will be saturated after three and half years of operation.
An event hub with two throughput units charges 0.06\$/hour and 0.03\$/million messages, incurring an hourly charge of approx. 7\$ for the 
The hourly rate for the data explorer for the given specifications is 1.5\$ for the \textit{pay as you go} tier.
\textbf{The first use case requires 48.5\$/hour, with Azure DT taking up most of the cost}. 

The lag of the location parameter in TwinRAN was measured for the second use case using the method explained in Section\ref{sec:evaluating_adt}. 
The update from the source first acts on its twin, then traverses the event hub and is routed to TwinRAN. 
With 10,000 measurements, the average lag for this setup was found to be 140 ms, with the maximum being 562 ms. 
If we update the location of the UE 5 times a second, this lag should still allow the latest location information at the TwinRAN instance. 
However, an object moving at 200 km/hr would have moved roughly 10 meters in 200 ms. Therefore, the high latency of the Twin-Twin update would restrict the second use case to slowly moving objects such as AGVs.  
The cost of running the \textbf{second use case is 47\$/hour} - the same as the first minus the 1.5\$ for the data explorer cluster. 

We spawned five copies of TwinRAN using a custom-made Azure function 100 times. 
We timed the function from when it received the trigger to when the last instance copy update was successful. 
This spawning time was 22 seconds on average, with a maximum of 34 seconds. 
Azure DTs charge for a query based on a metric called query unit. 
The querying of one TwinRAN deployment of our network size takes 800 querying units. 
Azure DTs charge 0.50\$ per million query units. 
The Azure function makes 107 executions for spawning the 5 twin copies. 
Azure DTs and functions charge 0.50\$ per million query units and 0.20\$ per million function executions, and therefore, these costs would be much less compared to the 40\$/hour for each TwinRAN copy that would be spawned. 
\begin{table}[]
\resizebox{\columnwidth}{!}{%
\begin{tabular}{|c|c|c|c|}
\hline
\textbf{Use Case} & \textbf{KPI Name}          & \textbf{KPI Value} & \textbf{Cost of Running} \\ \hline
1                 & Data Volume                & 48MB/hour          & 48.5\$/hour              \\ \hline
2                 & Twin-Twin Update Latency   & 140 ms             & 47\$/hour                 \\ \hline
3                 & Spawning Time for 5 copies & 22 s               & 40\$/hour/copy             \\ \hline
\end{tabular}%
}
\caption{\small{Results showing each use case's selected KPI and pricing. Most of the cost comes from running the Azure DT at full capacity in the cloud at 40\$ per hour. Adding applications (use cases) is relatively cheaper, as seen from the first and second rows. Hence, scaling TwinRAN to more use cases can increase cost efficiency. This scaling is easier due to the non-invasive, multipurpose, and vendor-agnostic characteristics of TwinRAN.}}
\label{tab:results}
\end{table}

\section{Limitations and Conclusion}
\label{sec:conclusion}
In this paper, we presented a concept and architecture for twinning the 5G RAN on the cloud using Azure DT.
The architecture focuses on achieving three main properties of TwinRAN: vendor-agnostic, multipurpose, and non-invasive.
TwinRAN incorporates and utilizes the O-RAN concept, seamlessly integrating it into a 5G RAN.
Thorough investigations of the behavior of Azure DTs help us make design decisions for TwinRAN, such as maintaining two different instances simultaneously for intercell and intracell operations. 
The investigations also show some limitations - it is not a suitable solution for tightly coupled close-loop data-driven operations requiring latencies smaller than 100 ms, such as ChARM\cite{charm} for spectrum sharing.
We demonstrate three use cases where TwinRAN is used, showing its multipurpose nature. 
We analyze the costs of running TwinRAN under the currently openly available cloud pricing terms. 
Analysis shows that the operating costs are dominated by the prices of maintaining the DT in the cloud at \textbf{0.05\$ per user per hour} while the use cases themselves do not add significantly to it.
These costs currently make it untenable for operators to deploy TwinRAN all the time over their entire network. 
Therefore, with the current pricing structure, TwinRAN might be beneficial to be deployed for a partial part of an operator's network for a limited amount of time to tackle specific problems like heavy traffic, what-if studies, or collecting data for training or publishing. 
The same TwinRAN instance can execute all three use cases simultaneously, and costs do not increase as more applications are deployed. 
 We believe that if an operator were to implement TwinRAN, they could negotiate with Azure for a subscription at better pricing and scale it to many more instances.
 In this work, we are limited to 100 twins/instance due to the update rate, as mentioned in Section \ref{subsec:model_sizes_etc}, and to 8 instances due to the tier-level subscription to Azure DT provided by our academic institution.
 
Our contribution underscores the feasibility and advantages of DT technology in the context of 5G networks. 
It sets the stage for future network management and optimization innovations using cloud-based digital twins. 
Our architecture aims to facilitate more intelligent, resilient, and adaptive 5G networks by bridging the gap between physical and digital network environments.

\bibliographystyle{IEEEtran}
\bibliography{references,new_references}

\end{document}